\documentclass[%
 reprint,
nofootinbib,
 amsmath,amssymb,
 pre,
]{revtex4-2}
\usepackage{graphicx}
\usepackage{float}

\begin{document}
\title{Generalized end-product feedback circuit senses high dimensional environmental fluctuations}

\author{Fang Yu}
\author{Mikhail Tikhonov}
\date{May 2022}

\begin{abstract}
   Understanding computational capabilities of simple biological circuits, such as the regulatory circuits of single-cell organisms, remains an active area of research. Recent theoretical work has shown that a simple regulatory architecture based on end-product inhibition can exhibit predictive behavior by learning fluctuation statistics of one or two environmental parameters. Here we extend this analysis to higher dimensions. We show that as the number of inputs increases, a generalized version of the circuit can learn not only the dominant direction of fluctuations, as shown previously, but also the subdominant fluctuation modes.
\end{abstract}

\maketitle

\section{Introduction}
As organisms evolve to better survive in changing environments, they develop adaptations that allow them to respond to change, but also to predict change. Characterizing such predictive (anticipatory \cite{Tagkopoulos, Savageau}) behavior in microorganisms, whose regulatory architectures are far less complex than can be achieved by a neuron-based brain, revealed many examples of evolutionary ingenuity attaining complex objectives with minimal ingredients (e.g., robust circadian clocks in photosynthetic algae, which allow them to reorganize their metabolism in preparation for sunrise~\cite{Pedersen}). 

Theoretical computer science has long established that even the simplest building blocks, if used in sufficient numbers, can support complex computations: very simple instruction sets can already be Turing-complete~\cite{Turing, Brainerd, Perez}. The biologically relevant sister question---how simple of a \emph{circuit} can perform how complex a task?---is understood less well. Some well-studied examples include the chemotaxis circuit achieving perfect adaptation \cite{Yi, Chang}, mechanisms of temperature compensation in circadian clocks \cite{Pedersen, Husain}, or the bistable genetic regulatory network storing and retrieving associative memories \cite{Sorek}. Still, understanding the computational capabilities of biological circuits remains an active area of research. 

Recent theoretical work explored the ability of a simple circuit to learn statistical features of fluctuating environmental parameters \cite{Stefan}. It was shown that this task can be solved (in fact, near-optimally) by a simple implementation based on the end-product inhibition motif. That work considered the proof-of-principle cases of one or two environmental parameters, sufficient to demonstrate how the circuit can learn, respectively, the variances and correlations of its inputs and to discuss implementations with biologically plausible ingredients. However, optimization tasks solved by cells (e.g.\ resource allocation) are often high-dimensional. It is natural to ask whether the same architecture would retain its performance when the number of fluctuating factors increases. 

Here, we explore a high-dimensional generalization of the circuit proposed in Ref.~\cite{Stefan}. We ask whether the three ingredients identified in that work---nonlinearity, an excess of regulators, and cross-talk between them---are sufficient to learn the fluctuation structure of high-dimensional environments. 

We find that, as in the low-dimensional case, the circuit can upregulate its reactivity to respond faster in epochs when environment fluctuations are larger. In principle, this allows it to outperform simple end-product inhibition at the tracking task defined in \cite{Stefan}, but we argue that in the high-dimensional case this small performance gain is unlikely to justify the significant complexity cost. However, we further show that even a small excess of regulators already makes the circuit responsive to changes in fluctuation structure, and that the state adopted by the circuit encodes both the dominant and subdominant fluctuation modes of environmental parameters. We conclude that the generalized end-product feedback circuit could serve as a sensor of subtle environmental change.

\section{The model}
Our approach builds directly on that of Landmann et al.~\cite{Stefan}, but this section provides enough details to be self-contained.

Specific adaptation problems faced by real organisms are highly diverse. Following Ref.~\cite{Stefan}, here we distil the general problem of physiological learning to a minimal model. Specifically, we consider a scenario where a set of internal quantities $\vec{P} = (P_1,\dots,P_N)$ (which a cell can regulate) must track a set of fluctuating external factors, $\vec{D} = (D_1,\dots,D_N)$. For the sake of concreteness, we will think of this problem in metabolic terms, with $P_i$ representing the rates of production of metabolites $x_i$. In our model, the cell seeks to match these production rates $\vec P$ to the (time-dependent) demands $\vec{D}(t)$ imposed by the external conditions. As an example, environmental conditions that trigger biofilm formation in bacteria require a different stoichiometry of synthesis than the condition of fast planktonic growth. 

If the fluctuations of demands are slow, the organism could sense them and directly match $\vec P$ to $\vec D$ at all times. But if fluctuations are too fast to be followed precisely, the organism must instead rely on the ``statistical structure'' of $\vec D(t)$, such as the mean value or correlations between its components $D_i(t)$~\cite{Stefan}. If this statistical structure remains constant over a very long timescale, the optimal behavior (given this structure) could be hardwired into the circuit by evolution. But if the structure itself occasionally changes, the organism would need to learn it from recent observations via physiological mechanisms. This is the regime where this problem can serve as a minimal model for the task of physiological learning.


There are different levels of statistical structure to be learned. Under our tracking problem, the simplest form of learning would be to set the production rates $P_i$ to match the average demand in the recent past. Beyond that, the subtler statistics include the variances and correlations among fluctuations. To model $D(t)$ in a way where both means and correlations can be tuned, we consider a multi-dimensional random walk in a quadratic potential~\cite{Stefan}.

\begin{equation}
    \vec{D}(t+\Delta t) = \vec{D}(t) - M\Delta t\cdot(\vec{D}(t) - \vec{\langle D\rangle}) + \sqrt{2\Gamma\Delta t}\,\vec{\eta}\label{eq:1}
\end{equation}
Here $\vec{\langle D\rangle}$ denotes the average demand, $\Gamma$ denotes the fluctuation strength, and $\vec{\eta}$ is a series of independent Gaussian random variables with zero mean and unit variance. The matrix $M$ determines the correlation among fluctuations of different components of $D$. If $M$ is isotropic, the fluctuations of individual components of $D$ will be decoupled.

Our approach will be as follows. To probe whether a given regulatory architecture successfully learns the statistical structure of the fluctuating environment, we expose it in simulations to several environmental epochs that differ by statistical structure. To say that the system successfully ``learns'' its environment, we require two criteria. First, the system should be sensitive to the change of statistics, i.e.\ we expect the regulator activity to be reorganized between epochs. Second, we should be able to exhibit the ``rule'' by which the statistical feature of interest is encoded in regulator activity. 

\begin{figure}[h!]
\centering
  \includegraphics[width=0.4\textwidth]{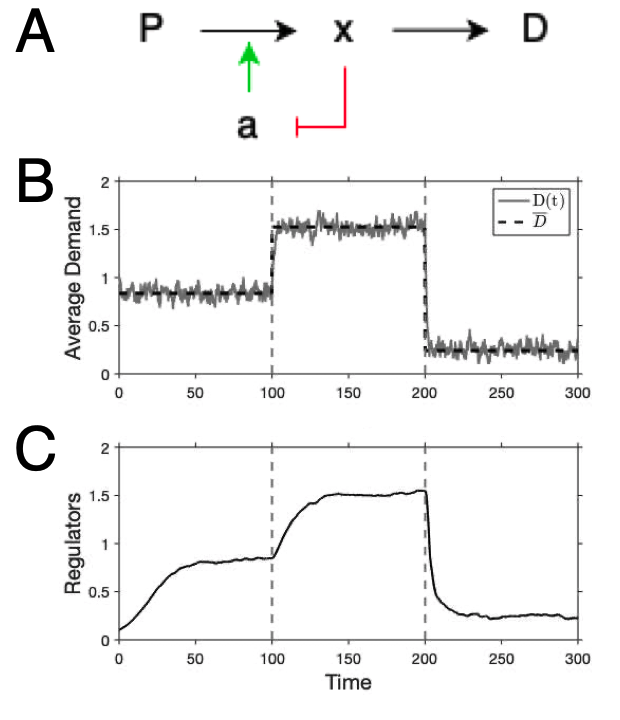}
  \caption{The SEPI architecture can be seen as learning signal mean. (A) Simple end-product inhibition (SEPI). (B) We expose the SEPI architecture to 3 environmental epochs (solid line) that differ by the average demand $\vec{\langle D\rangle}$ (dashed line). (C) The expression level of the regulator $a$ encodes the average demand. }
  \label{fig:1}
\end{figure}

The simplest form of statistical structure is the average demand $\vec{\langle D\rangle}$. This average demand can be learned already by the simple end-product inhibition (SEPI) circuit, where the production $P_i$ of each metabolite $x_i$ is placed under control of a single dedicated regulator $a_i$ inhibited by $x_i$ itself (Fig.~\ref{fig:1}A). In epochs of low demand, the unused $x_i$ accumulates and decreases production until it balances the demand. The average demand over a recent past is stored in the activity of the regulator $a_i$ (Fig.~\ref{fig:1}B, C).

The regulatory motif of end-product inhibition is not only simple, but has been shown to be remarkably effective. For example, under certain assumptions, this motif alone can not only ``solve'' the problem of proteome reallocation after a change of environmental conditions, but do so in an optimal time \cite{Pavlov}. However, the effectiveness of SEPI necessarily applies only when dealing with states that, in the language of our model, differ by the signal mean. The internal degrees of freedom (the regulators $a$, serving as memory) can store only one value per metabolite $x_i$. To be sensitive to additional statistics, additional degrees of freedom would necessarily be required. Thus, from now on, we will allow the number of regulators $N_a$ to exceed the number of metabolites $N_x$, and label regulators using Greek indices $\mu$, running from 1 to $N_a$ (while Roman indices $i$, labeling metabolites, run from 1 to $N_x$).

Landmann et al.\ showed that a generalized end-product feedback architecture can learn the variances and correlations of $D_i$~\cite{Stefan}. Their architecture takes three ingredients: an excess of regulators ($N_a > N_x$), nonlinear activation/repression of the regulators $a_\mu$ by the metabolite concentrations $x_i$, and cross-talk among different regulatory pathways. Specifically, they considered the following dynamics:

\begin{equation}
    \begin{aligned}
    x_i &= \frac{P_i}{D_i} \\
    P_i &= \sum_\mu\sigma_{\mu i}a_\mu \\
    \tau_a \dot{a}_\mu &= a_\mu \max\Big(d,\sum_i\sigma_{\mu i}(1 - x_i)\Big) - \kappa a_\mu \\
    \end{aligned}
    \label{eq:2}
\end{equation}
Here $\sigma_{\mu i}$ describes how the activities of regulators $a_\mu$ control the synthesis of metabolites $x_i$; $d$ parameterizes nonlinearity; and $\kappa$ is the degradation rate.

\begin{figure*}[t!]
  \includegraphics[width=0.95\textwidth]{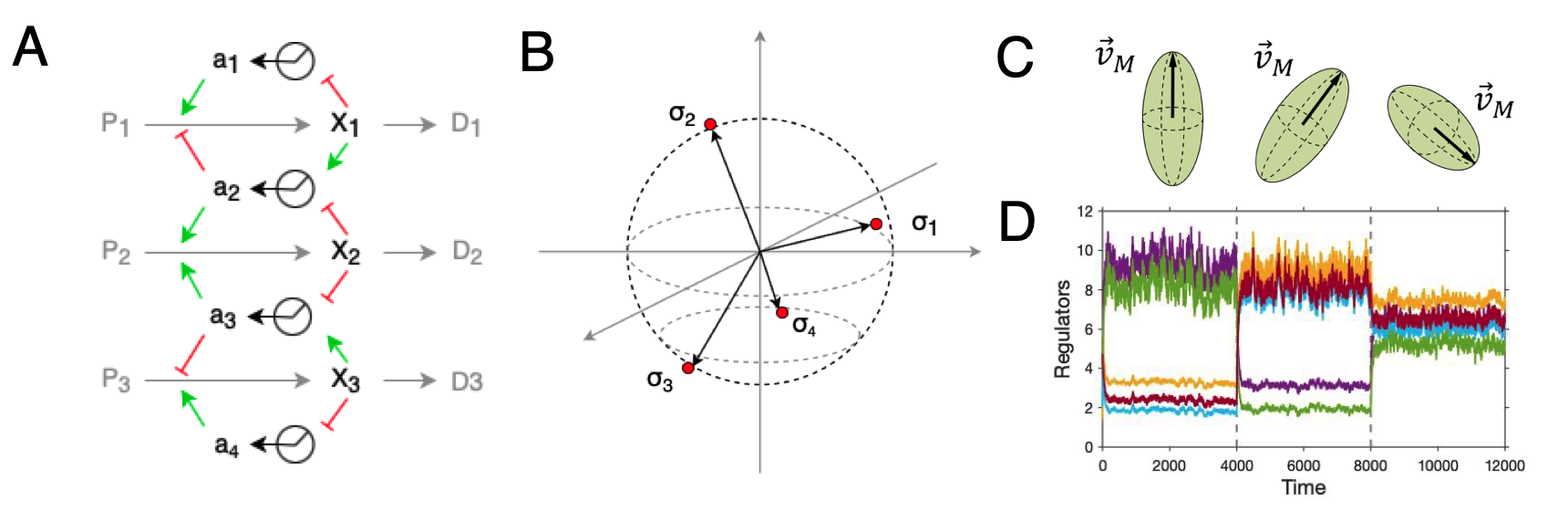}
  \caption{The generalized architecture is built to learn the variances and correlations of demand $\vec{D}$. (A) An example of generalized end-product feedback architecture in the case where $N_x=3$ and $N_a=4$. (B) 4 regulatory pathways on 3 resources are shown as 4 vectors in 3d space. Regulation vectors are simulated as charges repelling each other to distribute evenly. (C) To test the ability of the system to learn, we expose it to 3 environmental epochs overtime differed in the dominant eigenvector of $M$. shown here ellipsoids representing the orientation of the eigenbasis of $M$. (D) In response to these changes, the system dynamically adjusts the expression level of regulators. }
  \label{fig:2}
\end{figure*}

In two dimensions ($N_x=2$), this circuit can sense, store and usefully ``recall'' the information on second-order input  statistics, such as variances and correlations, and do so near-optimally~\cite{Stefan}. Here, we extend this architecture to higher dimensions (Fig.~\ref{fig:2}A). We choose regulators $\sigma$ to be minimally redundant (see SI section B). Briefly, the elements of $\sigma$, normalized as $|\sigma|=1$, can be seen as points on a sphere, and we pick them to be spread out as far away from each other as possible by treating them as repelling charges on a sphere's surface (Fig.~\ref{fig:2}B). We expose the generalized end-product feedback architecture to environmental epochs that differ in fluctuation structure $M$ only. To guarantee that any restructuring of regulator activity between exposure epochs is due to the changes of $M$, we keep the mean demand $\bar D$ the same in all epochs. For concreteness, we pick $M$ to be a random rotation of $\left(\begin{smallmatrix}
  1 & 0 & \cdots & 0\\
  0 & 100 & \cdots & 0\\
  \vdots & \vdots & \ddots & \vdots\\ 
  0 & 0 & \cdots & 100
\end{smallmatrix}\right)$, so that the environmental fluctuations have one preferred direction. When shifting from one epoch to another, we reorient $M$ by applying a random rotation (Fig.~\ref{fig:2}C), and observe how the regulatory architecture reorganizes the regulator expression levels in response (Fig. \ref{fig:2}D).

Note that changing the direction of the dominant eigenvector is only one way to change the environment. For example, Ref.~\cite{Stefan} also considered environments with different extent of correlation among fluctuations of $D_i$'s, which the circuit was also able to learn. Here, we will use the former approach, because the existence of a preferred direction of fluctuations allows for more visual metrics to quantify how well the circuit learned, as we describe below. We will show that the architecture of Fig.~\ref{fig:2}A is indeed responsive to higher-dimensional rotations of $M$, and will quantify this sensitivity. 
 
\section{Results}
\subsection{The generalized architecture\\ can outperform SEPI, but is costly}

When Landmann et al considered the architecture (Eq~\ref{eq:2}) in the two-dimensional case, their focus was not just learning, but also the benefit of learning. Specifically, their primary readout was the tracking performance defined by $-\sqrt{\sum_i(P_i-D_i)^2}$, a proxy for organism fitness, and they showed that the learning-capable circuit can enhance tracking performance over SEPI. In this section, we demonstrate that this observation continues to hold in higher dimensions: namely, the circuit of Fig.~\ref{fig:2}A can achieve better tracking performance than the SEPI architecture. However, we will also show that this performance increase is very costly.


Here we use two metrics of cost: Control input power (CIP), defined by $\int{||\dot{P}||^2dt}$, is a concept borrowed from control theory. Measuring cost in this way has the advantage that the family of optimal strategies on the performance-CIP plane can be derived analytically; but CIP is not easy to interpret in biological terms. A more biologically relevant measure of cost is the total expression of all regulators combined: $\sum_\mu a_\mu$. 


\begin{figure}[t!]
  \includegraphics[width=\linewidth]{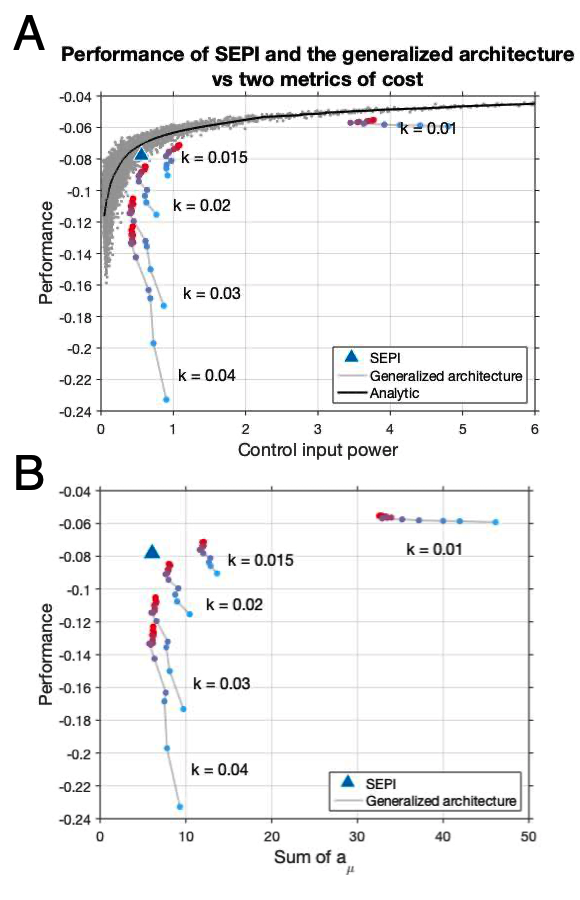}
  \caption{Improving performance beyond SEPI is very costly. (A) Tracking performance of different architectures when $N_x=6$, shown against Control Input Power (CIP), which is a measure of cost (see text). Red dots show simulation results of the optimal strategy, and the red line is the smoothed curve. The blue dot indicates SEPI performance, and the grey line shows the performance of our architecture, with dots changing from blue to red indicating an increasing number of regulators. (B) Same as A, replotted using a more biologically relevant measure of cost (the total expression of all regulators $\sum_\mu a_\mu$). }
  \label{fig:3}
\end{figure}

Fig.~\ref{fig:3} confirms that both cost metrics yield similar results. As expected, increasing the number of regulators increases performance. Performance can also be improved by reducing the degradation rate $k$, since a higher expression of regulators (with activators and repressors active simultaneously, known as paradoxical regulation~\cite{Hart}) allows $P_i$ to change faster~\cite{Stefan}. As a result, the performance of the generalized architecture can exceed that of SEPI. However, Fig.~\ref{fig:3} also shows the remarkable effectiveness of SEPI, which lies closest to the optimal curve at minimal circuit complexity. As described in Ref.~\cite{Stefan}, the advantage of the architecture (Eq.~\ref{eq:2}) is not to outperform SEPI in the absolute sense of the Pareto front of performance vs.\ cost. Rather, it provides a way of investing extra resources into improving performance at some important task. Such mechanisms are known to be employed by cells in other contexts, e.g. investing energy to improve the accuracy of sensing~\cite{Laughlin,Lan} or copying its DNA~\cite{Olson}. However, in our context it seems implausible that this marginal performance increase alone would be sufficient to offset the cost of a significant increase in protein expression levels and circuit complexity. Thus, from here on, we will no longer consider tracking performance as our readout. Instead, we will assume that an ability to sense subtle changes in environmental statistics may itself be of value to the organism (e.g.\ as an early cue indicative of some other upcoming change), and investigate the ability of this circuit to learn the environmental state and react to its changes.

\subsection{The architecture is sensitive to and tracks the dominant eigenvector of M}

\begin{figure}[b!]
  \includegraphics[width=\linewidth]{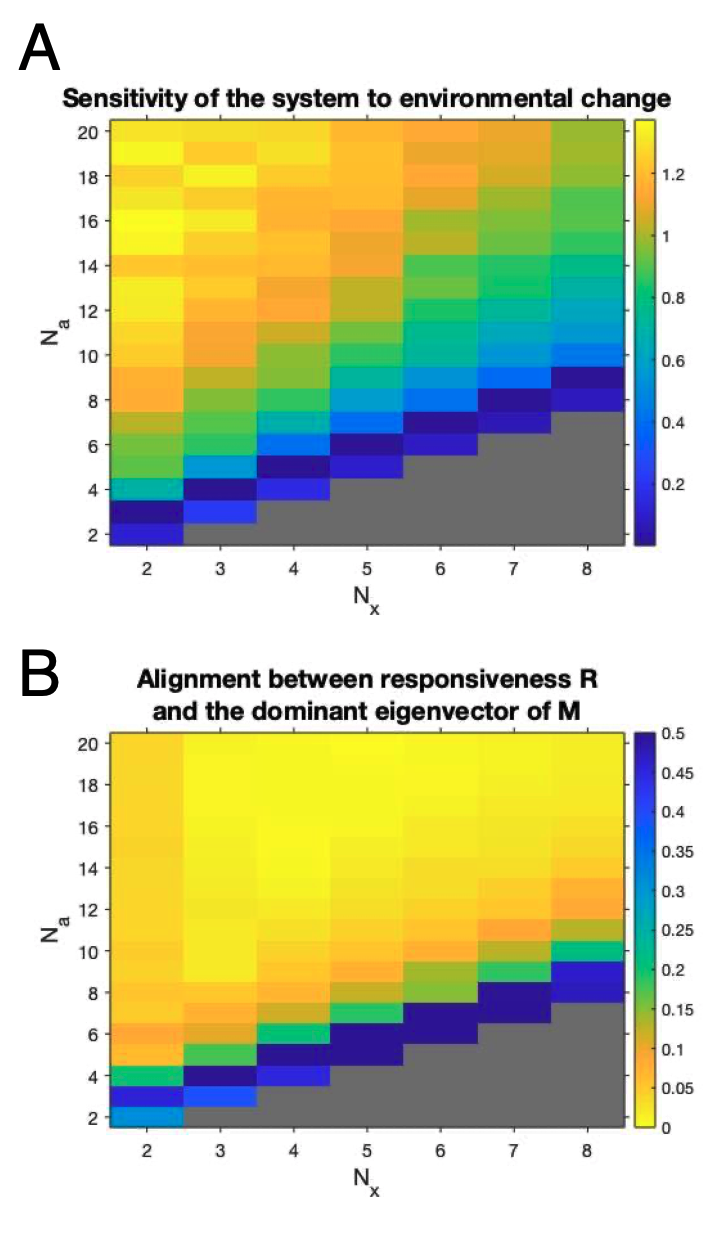}
  \caption{A modest excess of regulators allows the generalized architecture to learn. (A) Heatmap of sensitivity $\chi$ shows that the generalized architecture responds to environmental change with an excess of regulators. (B) Heatmap of the alignment $\gamma$ between responsiveness $R$ and the dominant eigenvector of $M$ indicates that with a modest excess of regulators, the generalized architecture tracks the dominant eigenvector of $M$ by aligning its responsiveness towards it. }
  \label{fig:4}
\end{figure}


To `learn' fluctuation structure, our architecture must first be sensitive to it. This means that when environmental statistics change, the expression level of regulators should change. To quantify this, we compare the expression levels before and after each environmental change, and define the sensitivity $\chi$  as the average angle between them  (interpreted as vectors in the $N_a$-dimensional expression space): 
\[
\chi=\frac 1{K-1}\sum_{k=1}^{K-1}\arccos \left[\frac{\vec{a}^{\,(k)}\cdot\vec{a}^{\,(k+1)}}{\|\vec{a}^{\,(k)}\|\,\|\vec{a}^{\,(k+1)}\|}\right].
\] Here $K=10$ is the number of environment epochs, and $\vec a^{\,(k)}$ denotes the average expression of all regulators in epoch $k$ (estimated by averaging the fluctuating expression over the second half of the epoch, when the expression reaches steady state; compare to Fig. 2D). Fig.~\ref{fig:4}A confirms that the architecture (Eq.~\ref{eq:2}) responds to the change in environmental statistics (Eq.~\ref{eq:1}), and is increasingly sensitive to it as the number of regulators increases.

We will now show that the regulation state adopted by the system is not random or idiosyncratic, but encodes information about $M$ in a simple way. Namely, we will show that the system preferentially aligns its responsiveness, $R_{ij} = \frac{d\dot{P}_i}{dD_j}$, to the dominant eigenvector of $M$. This behavior is, in fact, the ``smart'' thing to do: it can be shown that the optimal strategy (in the sense of control theory, with CIP as cost metric) would be similarly anisotropic, with the dominant direction of fluctuations eliciting the strongest response~\cite{Stefan, Liberzon}. To quantify this degree of alignment, we define \[
\gamma=\mathrm{Prob}_{\|u\|=1}\Big(\|R\cdot \vec{u}\|>\|R\cdot \vec{v}_M\|\Big)
\]
where $v_M$ denotes the dominant eigenvector of $M$. This metric quantifies the probability that a randomly drawn unitary vector $\vec{u}$ is aligned as good or better than $v_M$ to the responsiveness $R$. Unlike sensitivity $\chi$'s which are angles measured in increasingly high-dimensional spaces and are therefore affected by dimensional effects, $\gamma$ is defined to be comparable across dimensions. The smaller the value of $\gamma$, the stronger the evidence that our generalized end-product feedback architecture adopted a state with a responsiveness matrix preferentially aligned to $v_M$.
Heatmap of $\gamma$ (Fig.~\ref{fig:4}) shows that even just two excess regulators exhibit clear signature of the architecture aligning better than random.

\begin{figure}[t!]
  \includegraphics[width=0.45\textwidth]{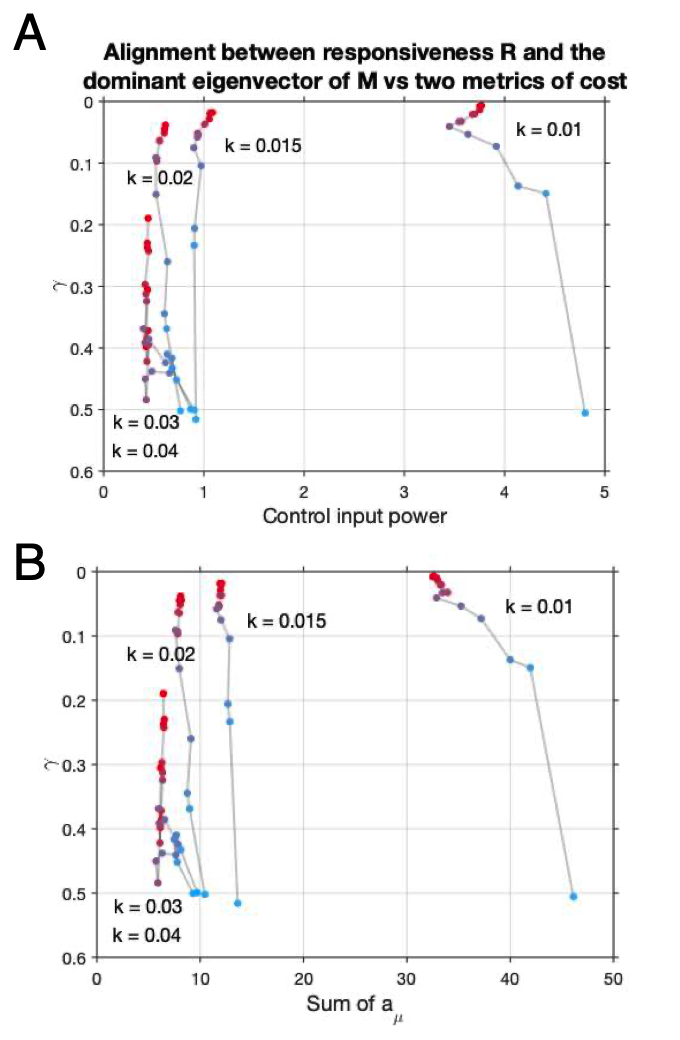}
  \caption{More regulators enhance learning of fluctuation structure without extra cost. (A) $\gamma$ (alignment between responsiveness $R$ and the dominant eigenvector of $M$) vs CIP in the case of $N_x = 6$ for different $N_a$ and $\kappa$. (B) $\gamma$ vs CIP in the same condition. The color of dots changing from blue to red indicates an increasing number of regulators.}
  \label{fig:5}
\end{figure}

Learning the fluctuation structure may also require some cost. To measure this, we show $\gamma$ vs CIP and the total expression level of regulators ($\sum a_\mu$) in Fig.~\ref{fig:5}.
In contrast to Fig.~\ref{fig:3}, Fig.~\ref{fig:5} shows that the learning ability of the circuit is controlled primarily by the number of regulators instead of degradation rate, and can be modulated without incurring an expression or CIP cost.

\subsection{The architecture also tracks non-dominant statistics}

\begin{figure*}[t!]
  \includegraphics[width=\linewidth]{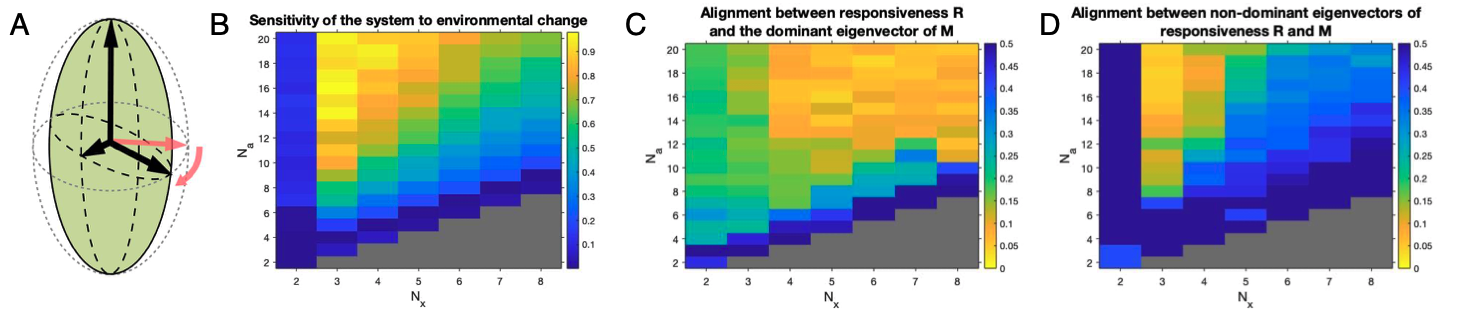}
  \caption{Our architecture learns to track the changing sub-dominant eigenvector of the input fluctuation. (A) We rotate $M$ with its dominant eigenvector fixed. (B) Heatmap of the rotation angle of vector $\vec{a} = \{a_1,\dots, a_\mu,\dots, a_{Na}\}$ indicates that our architecture is sensitive to changing subdominant eigenvector of $M$.  (C) Heatmap of the alignment $\gamma$ between responsiveness $R$ and the dominant eigenvector of $M$. (D) Heatmap of the alignment $\digamma$ between non-dominant eigenvectors of $R$ and $M$.}
  \label{fig:6}
\end{figure*}

Till now we have been considering the case where $M$ has only one dominant direction, but what happens when $M$ is more complex? To test this, we consider $M$ with a sub-dominant eigenvector---specifically, a randomly rotated version of $M_0=\left(\begin{smallmatrix}
  1 & 0 & 0 & \cdots & 0\\
  0 & 2 & 0 & \cdots & 0\\
  0 & 0 & 100 & \cdots & 0\\
  \vdots & \vdots & \vdots & \ddots & \vdots\\ 
  0 & 0 & 0 & \cdots & 100
\end{smallmatrix}\right)$---and restrict environment changes to those that change only the sub-dominant direction, keeping the dominant eigenvector fixed. For example, in a three-dimensional case, one can intuitively think of this as rotating an anisotropic ellipsoid around its dominant axis (Fig.~\ref{fig:6}A). Fig.~\ref{fig:6}B confirms that for any $N_x>2$, the system indeed alters its regulatory state in response to such environmental changes. The $N_x=2$ column serves as a control: in two dimensions, the only rotation that preserves the dominant eigenvector is the identity matrix. Thus, the environmental statistics do not change between epochs, and any observed differences in ``regulatory state'' are due to the noise in our estimation of average expression from finite simulation data (averaging the fluctuating expression over a finite simulation time). We observe that as long as $N_a\geq N_x+2$, the changes of regulatory state exceed the noise floor, confirming that the system is sensitive not only to the dominant direction of fluctuations (cf. Fig~\ref{fig:4}A), but to the direction of the second eigenvector as well. 

How well does the system learn? Plotting gamma indicates that the alignment to the dominant eigenvector (Fig.~\ref{fig:6}C) remains significant (better than random), but becomes worse than we observed in Fig.~\ref{fig:4}B. This is, of course, expected: the structure of fluctuations no longer reduces to a single dominant direction, so when assessing the alignment between the responsiveness $R$ and $M$, looking at only the dominant eigenvector of $M$ is insufficient. To also take non-dominant eigenvectors of $M$ into consideration, we defined the alignment $\phi$ between the whole matrices $R$ and $M$ as \[\phi_{[M,R]} = \frac{\|[M,R]\|}{\|\{M,R\}\|}\]


Where $\|\dots\|$ denotes the Frobenius norm of a matrix. If $M$ and $R$ are diagonal in the same basis, they commute and $\phi$ is zero. Thus, a non-zero $\phi$ can be seen as a measure of misalignment between the eigenbasis of $M$ and $R$.

Similar to the trick we did to $\gamma$, to enable meaningful comparisons across dimensions, instead of focusing on the raw value of $\phi$, we compute the probability \[\varphi=\mathrm{Prob}_{R'}(\phi_{[M,R']}<\phi_{[M,R]}).\]

Here $R'$ is a random rotation to $R$ with its dominant eigenvector fixed. If $\varphi$ is close to zero, it means that [...].



Fig.~\ref{fig:6}D shows that particularly in dimensions 3 and 4, the generalized architecture (Eq.~\ref{eq:2}) is not only responsive to changes in subdominant direction of fluctuations, but succeeds at realigning its responsiveness matrix accordingly.

\section{Discussion}


The regulatory circuit we considered in this work generalizes simple end-product inhibition by including three additional ingredients: nonlinearity, an excess of regulators, and cross-talk between them. Previous work has shown that these ingredients can endow the circuit with an ability to learn time-dependent fluctuation statistics of its inputs through a form of associative learning, at least in the low-dimensional scenarios (with one or two inputs)~\cite{Stefan}. Here, we generalized this circuit to the higher-dimensional case and presented two results. Just like in lower dimensions, this architecture can show an improved performance at the task of tracking environmental fluctuations; but we have argued that this small performance gain is unlikely to offset the significant complexity cost in practice. However, if sensing changes in environmental statistics is of value to the organism, then this architecture is quite interesting as it offers a sensitivity to subtle changes, sensing not only the dominant direction of fluctuations, but also the subdominant fluctuation modes. 

How relevant is this high-dimensional case for real cells? It is easy to imagine that a specific pair of resources might be correlated at some point of an organism's lifecycle but not at another; thus, it is clear that dynamically learn and unlearn a \emph{single} correlation could be useful. By comparison, the ``universal learning'' capacity described here---an ability to pick out any preferred direction of correlations in a high-dimensional space---seems rather more abstract. However, this could be the relevant regime for the regulatory architecture of a cell as a whole, seen as a high-dimensional learning circuit. Some intriguing recent ideas propose a possible common ground between evolvability of regulatory circuits, their ability to solve complex problems, and the success of overparameterized models in machine learning~\cite{Howell, Rocks}. So far, these parallels remain speculative; but the fact that simple elements can enable regulatory circuits to perform a form of associative learning could be a valuable  piece of this puzzle. In this work, we have shown that one previously proposed mechanism successfully generalizes to higher dimensions.








\clearpage
\section{Supplementary Information}
\subsection{Simple end-product inhibition}
When discussing learning the average demand $\vec{\langle D\rangle}$ in the model part, we introduced the simple end-product inhibition architecture. The specific dynamics of SEPI is given by the following equations (reproduced from Ref.~\cite{Stefan}):

\begin{equation}
    \begin{aligned}
    \dot{x}=P-D\frac{x}{x_0} \ &\mathrm{source-sink\ dynamics\ of\ metabolite\ x} \\
    P = aP_0 \ &\mathrm{definition\ of\ regulator\ activity\ a} \\
    \dot{a} = \frac{x_0 - x}{\lambda} \ &\mathrm{regulator\ activity\ inhibited\ by\ x}\\
    \end{aligned}
\end{equation}

Under the assumption that the dynamics of metabolite concentrations $x$ are faster than regulatory processes, and choosing the units so that $x$ and $P$ are dimensionless, SEPI dynamics becomes~\cite{Stefan}:

\begin{equation}
    \begin{aligned}
    x&=\frac{P}{D} \\
    P&=a \\
    \tau_a\dot{a}&=1-x \\
    \end{aligned}
    \label{eq:4}
\end{equation}

It is easy to see that SEPI learns the average demand over recent past within a time scale of $\tau_a$. Indeed, derived from Eq.~\ref{eq:4}, we get:
\begin{equation}
    \tau_a\dot{a}(t)=1-\frac{a(t)}{D(t)}
\end{equation}

Which gives
\begin{equation}
    \frac{1}{D(t)-a(t)}\mathop{dD(t)}-\mathop{d\ln(D(t)-a(t))}=\frac{\mathop{dt}}{\tau_aD}
    \label{SEPI:deduction}
\end{equation}

Noticing that $\langle\dot{D}(t)\rangle=0$. Therefore on average, the first term in Eq~\ref{SEPI:deduction} cancels. Integrating it over time leads to:

\begin{equation}
    \langle a(t)\rangle=\overline{D}-e^{-\frac{t}{\tau_a\overline{D}}}
    \label{SEPI:result}
\end{equation}

Where $\overline{D}$ is the average demand. The result of Eq.~\ref{SEPI:result} confrims that SEPI learns the average demand in a time scale of $\tau_a$.

\subsection{Charge repel simulation for distributing $\sigma_{\mu i}$}


Matrix $\sigma$ in Eq.~\ref{eq:2} represents regulatory pathways in a regulatory architecture. The $\mu$th row of it describes how the activity of regulator $a_\mu$ activates or inhibits various production $P$'s. 

We claim that regulatory pathways $\sigma_{\mu}$'s in higher dimensions need to be designed minimally redundantly. This requires $\sigma_\mu$'s to be scattered as dispersively as possible to reduce their relevance and explore the largest parameter space. As an example, if all $\sigma_\mu$'s point in the same direction, the performance will be poor. 

To achieve an overall good performance, we require $\sigma_\mu$'s to be distributed as far away from each other as possible. For example, in 1d case, the natural configuration is an activator and a repressor. In 2d, Landmann et al chose them as uniformly spaced points on a circle~\cite{Stefan}. Here, we generalize this to higher dimensions by simulating $\sigma_\mu$'s as point charges repelling each other in a $N_x$ dimensional space.


\end{document}